\documentclass[aps,twocolumn,showpacs,showkeys,amsmath,amssymb,superscriptaddress]{revtex4}
\usepackage{graphicx}
\usepackage{bm}      
\usepackage{color}
\newcommand{\lnco}{\textit{Ln}CoO$_3$}
\newcommand{\laco}{LaCoO$_3$}
\newcommand{\lasrco}{La$_{\rm 1-x}$Sr$_{\rm x}$CoO$_3$}

\newcommand{\prco}{PrCoO$_3$}
\newcommand{\ndco}{NdCoO$_3$}

\newcommand{\yco}{YCoO$_3$}

\newcommand{\LSiii}{LS Co$\rm ^{3+}$} 
\newcommand{\LSiv}{LS  Co$\rm ^{4+}$} 
\newcommand{\ISiii}{IS Co$\rm ^{3+}$} 
\newcommand{\HSii}{HS  Co$\rm ^{2+}$} %
\newcommand{\HSiii}{HS Co$\rm ^{3+}$} %
\newcommand{\teLSiii}{$t_{2g}^6e_g^0$}
\newcommand{\teLSiv}{ $t_{2g}^5e_g^0$}
\newcommand{\teISiii}{$t_{2g}^5e_g^1$}
\newcommand{\teHSii}{ $t_{2g}^5e_g^2$}
\newcommand{\teHSiii}{$t_{2g}^4e_g^2$}
\newcommand{\tsk}{$\rm \times 10^{-6} K^{-1}$}
\newcommand{\esplit}{E}
\begin{document}
\sloppy
\title{Structural anomalies, spin transitions and charge disproportionation in \textit{Ln}CoO$_3$
}
\author{Karel Kn\'{\i}\v{z}ek}
\email[corresponding author: ]{knizek@fzu.cz}
\author{Zden\v{e}k Jir\'{a}k}
\author{Ji\v{r}\'{\i} Hejtm\'{a}nek}
\affiliation{Institute of Physics ASCR, Cukrovarnick\'a 10, 162 53 Prague 6, Czech Republic.}
\author{Paul Henry}
\affiliation{Institut Laue Langevin, 6, rue Jules Horowitz, 38042 Grenoble Cedex 9, France.}
\author{Gilles Andr\'{e}}
\affiliation{Laboratoire L\'{e}on Brillouin, CEA-CNRS, CEA-Saclay, 91191 Gif-s-Yvette Cedex,
France.}
\begin{abstract}
The diamagnetic-paramagnetic and insulator-metal transitions in \lnco\ perovskites
(\textit{Ln}~=~La, Y, rare earths) are reinterpreted and modeled as a two-level excitation
process. In distinction to previous models, the present approach can be characterized as a
LS-HS-IS (low-high-intermediate spin) scenario. The first level is the local excitation of HS
Co$^{3+}$ species in the LS ground state. The second excitation is based on the interatomic
electron transfer between the LS/HS pairs, leading finally to a stabilization of the metallic
phase based on IS Co$^{3+}$. The model parameters have been quantified for \textit{Ln}~=~La,
Pr and Nd samples using the powder neutron diffraction on the thermal expansion of Co-O bonds,
that is associated with the two successive spin transitions. The same model is applied to
interpret the magnetic susceptibility of \laco\ and \yco.
\end{abstract}
\pacs{61.12.Ld; 75.30.Wx}
\keywords{\lnco, neutron diffraction, thermal expansion, spin-state transition.}
\maketitle
%


The perovskite cobaltites \lnco\ (\textit{Ln}~=~La, Y, rare-earth) systems show an insulating
ground state based on the diamagnetic low spin state of trivalent cobalt (LS, \teLSiii,
$S$~=~0). With increasing temperature the systems undergo two magnetic transitions connected
with excitations either to the intermediate spin (IS, \teISiii, $S$~=~1) or to the high spin
state (HS, \teHSiii, $S$~=~2).

Both transitions can be easily detected in the magnetic susceptibility in the case of parent
compound \laco. The first transition starts at about 50~K and saturated at about 150~K where a new
phase is stabilized with insulating character and Curie-Weiss susceptibility. The saturated
paramagnetic phase is manifested by a linear part of the inverse susceptibility over the region
150~-~350~K that gives $\mu_{eff}\sim$~3.2~$\mu_B$ and AFM interactions characterized by $\theta
\sim$~$-200$~K. Recent experiments at low-temperatures, including electron spin resonance (ESR)
\cite{RefNoguchi2002PRB66_94404}, inelastic neutron scattering \cite{RefPodlesnyak2006PRL97_247208}
and X-ray magnetic circular dichroism (XMCD) \cite{RefHaverkort2006PRL97_176405}, identify the
first transition as an excitation to HS state. The LS/HS scenario is also supported by band
structure calculations \cite{RefZhuang1998PRB57_10705,RefKnizek2006JPCM18_3285} that reveal three
possible ground states for \laco\ in the dependence of lattice volume and geometry: LS phase,
LS/HS(1:1) ordered phase and homogeneous IS phase. The calculation evidence that the HS state is
only stable if 6 nearest Co neighbors are in LS state, whereas stabilization of IS state is
promoted by IS neighbors. HS states prefers antiferromagnetic interactions, whereas IS states
prefers ferromagnetically aligned neighbors \cite{RefKnizek2006JPCM18_3285}.

The second transition of an insulator-metal (I-M) kind starts at about 500~K and is
accompanied with another increase of magnetic susceptibility and decrease of electric
resistivity to a metallic value $\rho\sim$~1~m$\Omega$cm. Above the I-M transition, another
region with quasilinear dependence of inverse susceptibility begins at about 600~K and spreads
up to 1000~-~1100~K. There is a clear change of slope with respect to region 150~-~350~K that
is traditionally interpreted as an increase of effective moments of Co$^{3+}$ while the
strength of AFM interactions is retained (see \textit{e.g.}
Ref.~\cite{RefHeikes1964PHYS30_1600}). We infer, however, that the observed susceptibility is
a combined effect of a change of AFM interactions towards FM ones and onset of temperature
independent Pauli paramagnetism, while the effective moments remain approximately the same.

This gives a possibility for a new interpretation of the spin transitions using the LS-HS-IS
scenario: The diamagnetic-paramagnetic transition in \laco\ originates in thermal population
of HS states in LS matrix at about 50~K. Each Co$^{3+}$ thus appears in a mixture of these two
states with increasing probability of the HS one that saturates finally in the 1:1 ratio at
about 150~K because of strong HS/LS nearest neighbor correlations. The second transition (I-M)
starts above 350~-~400 K and should be related to a charge disproportionation
\cite{RefKozhevnikov2003JSSC172_296,RefSehlin1995PRB52_11681}, in which $t_{2g}$ electron
transfer between LS/HS Co$^{3+}$ neighbors,
 \LSiii\ \teLSiii\ + \HSiii\ \teHSiii\ $\rightarrow$ \LSiv\ \teLSiv\ + \HSii\ \teHSii,
with charge transfer gap $\sim$~0.7~eV (see Fig.~5 in Ref.~\cite{RefKnizek2006JPCM18_3285}),
is an initial step. With increasing temperature, the probability of the electron exchange
increases and new IS Co$^{3+}$ states are gradually populated due to recombination reaction
 \LSiv\ \teLSiv\ + \HSii\ \teHSii\ $\rightarrow$ 2 \ISiii\ \teISiii.
This opens an $e_g$ conduction channel at about 500~K and leads finally to a formation of
metallic phase of IS character that coexists with residual regions of LS/HS kind with fast
dynamics. The optical conductivity data reveal indeed that this transition is associated with
a collapse of the $e_g$ charge transfer gap (originally $\sim$~2~eV) and the effect of high
temperature is thus analogous to the effects of heavy doping in \lasrco\ compounds
\cite{RefTokura1998PRB58_R1699}. The transitions temperatures for other \lnco\ compounds are
increasing with decreasing \textit{Ln} radius \cite{RefKnizek2005EPJB47_213}. In particular,
the onset of magnetic transition in \yco\ is shifted to 450~-~800~K and it develops
concurrently with the I-M transition without reaching the saturated LS/HS (1:1) phase.

In the present paper we apply the LS-HS-IS model to interpret the thermal expansion of Co-O
bonds of \lnco\ perovskites (\textit{Ln}~=~La, Pr and Nd) and quantify its parameters. The
transitions are treated in a rather formal way as a two-level excitation from LS/LS pairs to
LS/HS and IS/IS pairs. Since paramagnetic susceptibility is not sensitive to cobalt moments in
systems with magnetic rare earths, the present analysis is primarily based on observed
anomalies in thermal expansion that accompany both transitions due to the increasing ionic
radius of Co$^{3+}$ with increasing spin state
\cite{RefRadaelli2002PRB66_094408,RefZobel2002PRB66_R020402,RefBaier2005PRB71_14443,RefKnizek2005EPJB47_213}.


The thermal expansion data have been obtained by neutron diffraction experiments, performed at
low temperature in LLB (Saclay, France) on the G41 diffractometer using a wavelength
2.422~\AA\ and at high temperature in ILL (Grenoble, France) on the D20 diffractometer using a
wavelength 1.361~\AA. The observed neutron diffraction patterns were analyzed by a Rietveld
method with the help of the FULLPROF program (Version 3.30 - Jun2005-LLB JRC). The magnetic
susceptibility was measured on a SQUID magnetometer in the range up to 400~K using DC fields
10~kOe. The high temperature data up to 900~K were obtained using a compensated pendulum
system MANICS in a field of 10~-~19~kOe.


\begin{figure}
 \resizebox{0.85\columnwidth}{!}{\includegraphics[viewport=0 230 560 800,clip]{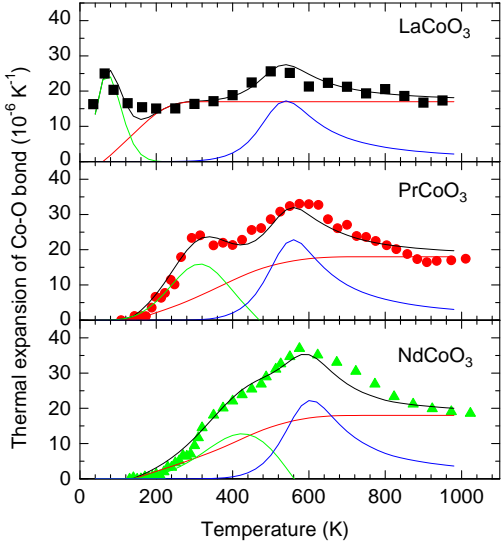}}
\caption{Experimental (symbols) and fitted (lines) thermal expansion of Co-O bond for \laco,
\prco\ and \ndco. The three contributions $\alpha^{latt}$, $\alpha_1^{mag}$ and
$\alpha_2^{mag}$ of the fitted thermal expansion are also displayed.} \label{FigLinExp}
\end{figure}

\begin{figure}
 \resizebox{0.85\columnwidth}{!}{\includegraphics[viewport=0 330 560 800,clip]{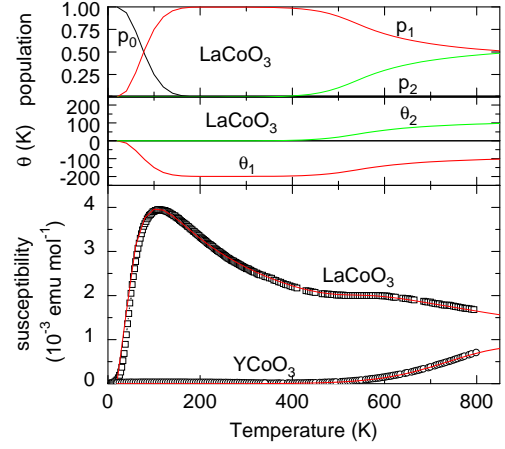}}
\caption{Upper: Population of ground and excited states. Middle: Evolution of Weiss $\theta$ for
excited states. Lower: Molar susceptibility for \laco\ and \yco. The solid line is fit based on our
model (see Eq.~\ref{EqCurieWeiss}).} \label{FigSus}
\end{figure}

The thermal expansion $\alpha$ of \lnco\ is a sum of three contributions: normal lattice
expansion $\alpha^{latt}$ and anomalous expansions due to the two spin transitions
$\alpha_i^{mag}$.

(1) Lattice term $\alpha^{latt}$ is a weighted sum by populations $p_i(T)$ of thermal expansions of
\lnco\ in individual magnetic states
\begin{equation}
 \alpha^{latt}(T) = \sum_{i=0}^{2} p_i(T) \; \alpha^{latt}_i(T)
\label{EqLinExpCalcLatt}
\end{equation}
where $i$~=~0 corresponds to ground state LS pairs, $i$~=~1 is the first excited state LS/HS
pairs and $i$~=~2 means the second excited state IS pairs. Lattice thermal expansion increases
with temperature and saturates around Debye temperature $\theta_D$. In our analysis we have
used the saturated values $\alpha^{latt}_0$~=~14\tsk,
$\alpha^{latt}_1$~=~$\alpha^{latt}_2$~=~18\tsk, and Debye temperatures $\theta_D \sim$~300~K
for \laco\ and $\theta_D \sim$~700~K for the other \lnco.

(2) The anomalous contribution $\alpha^{mag}$ can be expressed as
\begin{equation}
 \alpha_i^{mag}(T) =  \frac{\partial p_i(T)}{\partial T} \; d_i^{mag}(T)
\label{EqLinExpCalcMag}
\end{equation}
where $d_i^{mag}$ is related to the difference between the size of ground and excited spin
state and their dependence on temperature
\begin{equation}
 d_i^{mag}(T) =  d_i^{mag}(0) + \int_{0}^{T} \alpha_i^{latt}(T) - \alpha_0^{latt}(T) \; dT
\label{Eqdmag}
\end{equation}
In our fit we have found $d_1^{mag}(0)$~=~0.2\% and $d_2^{mag}(0)$~=~0.7\% for \laco\ and
$d_1^{mag}(0)$~=~0.3\% and $d_2^{mag}(0)$~=~1.0\% for the other \lnco.

Population of excited states $p_i$ ($i$~=1, 2) is calculated for each temperature point by
solving the set of 2 equations
\begin{equation}
 \frac{p_i(T)}{1 - \sum_{j=1}^2 p_j(T)} = \nu_i \; e^{-\esplit_i /T}
\label{EqPopulation}
\end{equation}
where $\esplit_i$ is the energy difference between the ground and excited states in units of
$T$, and $\nu_i$ is degeneracy of the excited state. The ground state has $\nu_0$~=~1. For the
first excited state $\nu_1$~=~3 can be anticipated for spin-orbit split state of HS Co$^{3+}$
that is isolated in diamagnetic matrix while the spin degeneracy $\nu_1 = 2S+1 = 5$ is more
appropriate for concentrated LS/HS pairs. For the second excitation $\nu_2 = 2\nu_1$ was used.

The energy difference for the first excited state $\esplit_1(T)$ depends on the structure changes
induced by temperature \cite{RefKnizek2006JPCM18_3285} or by pressure
\cite{RefVogt2003PRB67_R140401}. In our analysis the dependence of energy $\esplit_1(T)$ on
temperature was arbitrarily fitted by a power function \cite{RefKnizek2005EPJB47_213}
\begin{equation}
 \esplit_1(T) = \esplit_1^o \left[ 1 - \left( \frac{T}{T_o} \right)^n \right]
\label{EqEneFit1}
\end{equation}
where $\esplit_1^o$ is the energy splitting at $T$~=~0~K, $T_o$ is the temperature where $\esplit_1
(T_o)$~=~0 and $n$ is a fitting parameter that describes the curvature.

Since the stability of IS state is promoted by neighbors of the same kind, the energy
$\esplit_2(p)$ was set to depend on the concentration $p_2$ by the equation
\cite{RefKozhevnikov2003JSSC172_296,RefSehlin1995PRB52_11681}
\begin{equation}
 \esplit_2(p) = \esplit_1(T) + \esplit_2^o - \esplit_2^p \; p_2^{1/3}
\label{EqEneFit2}
\end{equation}
where $\esplit_2^o$ is the energy splitting for $p$~=~0 and $\esplit_2^p$ describes the dependence
on $p_2$.

The experimental and calculated thermal expansion of Co-O bond for \prco\ and \ndco, completed
with data on \laco\ taken from Ref.~\cite{RefRadaelli2002PRB66_094408}, are displayed in
Fig.~\ref{FigLinExp}.

\begin{table}
\caption{Ionic radii $r_{Ln}$ for 9-fold coordination and the parameters of the thermal
expansion fit for \textit{Ln}~=~La, Pr and Nd (see Eq.~\ref{EqEneFit1} and \ref{EqEneFit2} for
description of the parameters) and of the susceptibility fit for \textit{Ln}~=~La and Y.}
\begin{tabular*}{\columnwidth}{@{\extracolsep{\fill}} l|crrrrr}
\hline
        &  $r_{Ln}$ &  $E_1^o$ &  $T_o$ &   $n$ &  $E_2^o$ &  $E_2^p$ \\
\hline
        &     (\AA) &      (K) &    (K) &       &      (K) &      (K) \\
\hline
\laco\  &     1.216 &      160 &    100 &  3.00 &    2 600 &    2 500 \\
\prco\  &     1.179 &      890 &    370 &  3.11 &    2 700 &    2 600 \\
\ndco\  &     1.163 &    1 200 &    500 &  3.17 &    2 800 &    2 650 \\
\yco\   &     1.075 &    2 875 &    860 &  3.50 &    3 600 &    3 300 \\
\hline
\end{tabular*}
\label{Tab1}
\end{table}

The susceptibility was fitted by an equation
\begin{equation}
 \chi(T) = \frac{N_A \mu_B^2}{3 k_B}
           \sum_{i=1}^2 \frac{\mu_{i}^2 p_i(T)}{T -\theta_i(T)} 
\label{EqCurieWeiss}
\end{equation}
where $N_A$ is the Avogadro number,  $\mu_B$ the Bohr magneton, $k_B$ the Boltzmann constant
and $\mu_i$ is the effective moment of corresponding excited state. The Weiss temperature
$\theta$ was set to depend on the population of excited states $\theta_i(T) = \theta_i^o
p_i(T)$. In our fit we have found $\theta_1^o=-200$~K, $\theta_2^o=+200$~K, $\mu_1=\mu_2=3.25$
for \laco\ and 2.93 for \yco.

The temperature dependence of $p_i$  and $\theta_i$ for \laco\ and the experimental and
calculated molar susceptibility for \laco\ and \yco\ are shown in Fig.~\ref{FigSus}. The
parameters of the thermal expansion fit (La, Pr, Nd) and of the susceptibility fit (La, Y) are
summarized in Table~\ref{Tab1}.


In conclusion, the diamagnetic-paramagnetic and insulator-metal transitions in \lnco\ can be
successfully explained within a model of two-level excitation. The first level is the
excitation of HS Co$^{3+}$ species in the LS matrix. The steep temperature decrease of the
excitation energy close to the LS-HS crossover ($T_o$) is a signature that the local
excitations lead finally to a global phase transition. The second excitation is based on the
interatomic electron transfer and stabilization of IS Co$^{3+}$ states. The present
interpretation can be characterized as a LS-HS-IS scenario that is distinct from LS-IS-HS (or
LS-LS/HS-IS/HS) models previously used.

This work was supported by the Project No.~202/06/0051 of the Grant Agency of the Czech
Republic. The Institut Laue Langevin (Grenoble, France) and the Laboratoire Leon Brillouin
(Saclay, France) are thanked for providing access to the neutron beams and for all technical
support during the experiments.


\end{document}